# A test of the nature of cosmic acceleration using galaxy redshift distortions

L. Guzzo[1,2,3,4], M. Pierleoni[3], B. Meneux[5], E. Branchini[6], O. Le Fèvre[7], C. Marinoni[8], B. Garilli[5], J. Blaizot[3], G. De Lucia[3], A. Pollo[7,9], H. J. McCracken[10,11], D. Bottini[5], V. Le Brun[7], D. Maccagni[5], J. P. Picat[12], R. Scaramella[13,14], M. Scodeggio[5], L. Tresse[7], G. Vettolani[13], A. Zanichelli[13], C. Adami[7], S. Arnouts[7], S. Bardelli[15], M. Bolzonella[15], A. Bongiorno[16], A. Cappi[15], S. Charlot[10], P. Ciliegi[15], T. Contini[12], O. Cucciati[1,17], S. de la Torre[7], K. Dolag[3], S. Foucaud[18], P. Franzetti[5], I. Gavignaud[19], O. Ilbert[20], A. Iovino[1], F. Lamareille[15], B. Marano[16], A. Mazure[7], P. Memeo[5], R. Merighi[15], L. Moscardini[16,21], S. Paltani[22,23], R. Pellò[12], E. Perez-Montero[12], L. Pozzetti[15], M. Radovich[24], D. Vergani[5], G. Zamorani[15] & E. Zucca[15]

[1]INAF–Osservatorio Astronomico di Brera, Via Bianchi 46, I-23807 Merate (LC), Italy. [2]Max Planck Institut fur extraterrestrische Physik, [3]Max Planck Institut fur Astrophysik, [4]European Southern Observatory, D-85748 Garching, Germany. [5]INAF–IASF, Via Bassini 15, I-20133, Milano, Italy. [6]Dipartimento di Fisica, Universitá Roma III, Via della Vasca Navale 84, I-00146 Roma, Italy. [7]Laboratoire d'Astrophysique de Marseille, CNRS-Université de Provence, BP8, F-13376 Marseille cedex 12, France. [8]Centre de Physique Theorique, UMR 6207 CNRS-Université de Provence, F-13288 Marseille, France. [9]Astronomical Observatory of the Jagiellonian University, ul Orla 171, 30-244 Krakow, Poland. [10]Institut d'Astrophysique de Paris, UMR 7095, 98 bis Bvd Arago, [11]Observatoire de Paris, LERMA, 61 Avenue de l'Observatoire, F-75014 Paris, France. [12]Laboratoire d'Astrophysique de l'Observatoire Midi-Pyrénées (UMR 5572), 14 avenue E. Belin, F-31400 Toulouse, France. [13]INAF–IRA, Via Gobetti 101, I-40129 Bologna, Italy. [14]INAF–Osservatorio Astronomico di Roma, Via di Frascati 33, I-00040 Monte Porzio Catone, Italy. [15]INAF–Osservatorio Astronomico di Bologna, Via Ranzani 1, I-40127 Bologna, Italy. [16]Università di Bologna, Dipartimento di Astronomia, Via Ranzani 1, I-40127 Bologna, Italy. [17]Dipartimento di Fisica–Universitá di Milano-Bicocca, Piazza delle Scienze 3, I-20126 Milano, Italy. [18]School of Physics and Astronomy, University of Nottingham, University Park, Nottingham NG72RD, UK. [19]Astrophysikalisches Institut Potsdam, An der Sternwarte 16, D-14482 Potsdam, Germany. [20]Institute for Astronomy, University of Hawaii, 2680 Woodlawn Drive, Honolulu, Hawaii 96822, USA. [21]INFN–Sezione di Bologna, viale Berti-Pichat 6/2, I-40127 Bologna, Italy. [22]Geneva Observatory, ch. des Maillettes 51, CH-1290 Sauverny, Switzerland. [23]Integral Science Data Centre, ch. d'Ecogia 16, CH-1290 Versoix, Switzerland. [24]INAF–Osservatorio Astronomico di Capodimonte, Via Moiariello 16 I-80131, Napoli, Italy.

**Observations of distant supernovae indicate that the Universe is now in a phase of accelerated expansion[1,2] the physical cause of which is a mystery[3]. Formally, this requires the inclusion of a term acting as a negative pressure in the equations of cosmic expansion, accounting for about 75 per cent of the total energy density in the Universe. The simplest option for this 'dark energy' corresponds to a 'cosmological constant', perhaps related to the quantum vacuum energy. Physically viable alternatives invoke either the presence of a scalar field with an evolving equation of state, or extensions of general relativity involving higher-order curvature terms or extra dimensions[4-8]. Although they produce similar expansion rates, different models predict measurable differences in the growth rate of large-scale structure with cosmic time[9]. A fingerprint of this growth is provided by coherent galaxy motions, which introduce a radial anisotropy in the clustering pattern reconstructed by galaxy redshift surveys[10]. Here we report a measurement of this effect at a redshift of 0.8. Using a new survey of more than 10,000 faint galaxies[11,12], we measure the anisotropy parameter $b = 0.70 \pm 0.26$, which corresponds to a growth rate of structure at that time of $f = 0.91 \pm 0.36$. This is consistent with the standard cosmological-constant model with low matter density and flat geometry, although the error bars are still too large to distinguish among alternative origins for the accelerated expansion. This could be achieved with a further factor-of-ten increase in the sampled volume at similar redshift.**

A relevant consequence of the presence of a dominant form of dark energy in the Universe, in addition to its primary effect on the expansion rate, is to modify the gravitational assembly of matter from which the observed large-scale structure originated. In linear perturbation theory, it is possible to describe the growth of a generic small-amplitude density fluctuation through a second-order differential equation. This equation depends on the expansion rate $H(z)$, but also on the theory of gravity. From its solutions, we can define a linear growth rate $f$ that measures how rapidly structure is being assembled in the Universe as a function of cosmic time, or, equivalently, of the redshift. The redshift $z = \lambda_{meas}/\lambda_{emis} - 1$ of the radiation emitted by a distant object is a measure of the time of emission through its dependence on the cosmic scale factor $a(t)$, which is $1 + z = 1/a(t_{emis})$. $f(z)$ essentially depends on the value of the mass density parameter at the given epoch, $\Omega_m(z)$, which is defined as the ratio of the matter density $\langle\rho(z)\rangle$ to the 'critical' density required to halt the expansion $\rho_c = 3H(z)^2/8\pi G$, where G is Newton's constant. For the cosmological-constant model (in which the total density in matter and dark energy is $\Omega_m + \Omega_\Lambda = 1$) the dependence[9] is $f(z) \approx \left[\Omega_m(z)\right]^{0.55}$. However, this is not valid if the observed acceleration originates from a modification of the equations of the general theory of relativity; for example, in the Dvali-Gabadadze-Porrati (DGP) braneworld theory, an extra-dimensional modification of gravity[13], $f(z) \approx \left[\Omega_m(z)\right]^{0.68}$. In general, a fitting form $f(z) \simeq \left[\Omega_m(z)\right]^\gamma$ has been shown to be an accurate description for a wide range of models[9,14] (for which $\Omega_m(z)$ itself, not only $\gamma$, depends on the model). Thus, models with the same expansion history $H(z)$ but a different gravity theory will have a different growth rate evolution $f(z)$ and index $\gamma$ (refs 9, 15). A discrepancy between the measured value of the growth rate and that computed independently (assuming the general theory of relativity applies) from the $H(z)$ yielded by type Ia supernovae would point to modifications of gravity[6-8], rather than to exotic new ingredients in the physical content of the Universe[4,5].

A few observational techniques have been suggested to measure $f(z)$ at different redshifts[9,16]. Redshift-space distortions, that is, the imprint of large-scale peculiar velocities on observed galaxy maps, have not yet been considered in this context. Gravity-driven coherent motions are in fact a direct consequence of the growth of structure. The anisotropy they induce in the observed galaxy clustering when redshifts are used as a measure of galaxy distances can be quantified by means of the redshift-space two-point correlation function $\xi(r_p,\pi)$. Here, $r_p$ and $\pi$ are respectively the transverse and line-of-sight components of galaxy separations[17] (see Supplementary Information for definitions). The anisotropy of $\xi(r_p,\pi)$ has a characteristic shape at large $r_p$ that depends on the parameter[18,19] $\beta = f/b_L$. In practice, we observe a compression that is proportional to the growth rate, weighted by the factor $b_L$, the linear bias parameter of the specific class of galaxies being analysed. $b_L$ measures how closely galaxies trace the mass density field, and is quantified by the ratio of the root-mean-square fluctuations in the galaxy and mass distributions on linear scales[20]. Using this technique, a value of $\beta = 0.49 \pm 0.09$ has been measured at $z \approx 0.15$ using the 2dF Galaxy Redshift Survey (2dFGRS) sample of 220,000 galaxies with bias[10,21] $b_L = 1.0 \pm 0.1$, corresponding to a growth rate of $f = 0.49 \pm 0.14$.

We have measured the parameter $\beta$ at an effective redshift $z = 0.77$, using new spectroscopic data from the Wide part of the VIMOS-VLT Deep Survey (VVDS)[11,12]. The redshift-space correlation function $\xi(r_p,\pi)$ has been estimated from a recently completed subset of 5,895 faint galaxy redshifts between $z = 0.6$ and $z = 1.2$, covering an area of 4 square degrees (the F22 field; see Supplementary Information for more details). This corresponds to an effective sampling volume of $6.35 \times 10^6 h^{-3}$ Mpc$^3$, at a median epoch of ~7 Gyr, that is, about half the age of the Universe. $\xi(r_p,\pi)$ has been estimated in the conventional way by comparing the number of galaxy pairs at different separations $(r_p,\pi)$ to that in a random sample with an identical geometry and sampling pattern (Fig. 1a). The evident ellipsoidal shape (that is, the compression of the iso-correlation contours along the line of sight—the vertical direction) is the fingerprint of galaxy streaming motions. The corresponding value of $\beta$ can be measured by expanding the observed $\xi(r_p,\pi)$ in spherical harmonics; the coefficients of the expansion can be theoretically expressed as functions of $\beta$ and the best value of this parameter obtained through different fitting techniques[18,19]. We have directly tested these methods on fully realistic simulations of our data (M.P. *et al.*, manuscript in preparation); we obtained the least biased and most stable results through a direct maximum-likelihood fit of the full spherical harmonic model for $\xi(r_p,\pi)$, convolved with an exponential function that accounts for the small-scale nonlinear contribution[21] (see Supplementary Information for details). This model is characterized by two free parameters, the linear compression $\beta$ and the root-mean-square velocity dispersion of galaxy pairs $\sigma_{12}$, describing the small-scale incoherent motions in groups and clusters. The model that maximizes the likelihood, given our data, has $\beta = 0.70$ and $\sigma_{12} = 412$ km s$^{-1}$ (corresponding to the superimposed contours in Fig. 1a).

To estimate realistic errors for these values, we applied the same procedure to 100 independent mock replicas of our survey constructed from the Millennium simulation[22] including the full observing strategy, selection mask and redshift errors of the VVDS. These state-of-the-art simulations are highly successful in reproducing a wide range of galaxy and large-scale structure properties and can be considered in many respects as Monte Carlo realizations of our data. They allow us to include in the error budget a fair estimate of the finite sampling noise and of the 'cosmic variance' due to fluctuations on scales larger than the sampled volume. In Fig. 1b the contours correspond to the bivariate gaussian describing the distribution of the 100 mock measurements, centred on the best-fit $(\beta, \sigma_{12})$ pair from the data. The mean values of both parameters from the 100 mock catalogues $(\beta = 0.62 \pm 0.03, \quad \sigma_{12} = 382 \pm 12 \text{ km s}^{-1})$ are remarkably close to those measured from the data. This adds to our confidence in the overall realism of the simulations and consequently in the various tests performed to assess the robustness of our result (see Supplementary Information for details). Marginalizing over the root-mean-square pairwise dispersion (that is, integrating along $\sigma_{12}$), we obtain an estimate of the error on the compression parameter, such that $\beta = 0.70 \pm 0.26$.

Because both the growth rate and galaxy bias evolve with redshift, this value represents a mean over the redshift range $0.6 < z < 1.2$, weighted by the radial selection function of the sample. We take the effective redshift for this measurement to be $z = 0.77$, which corresponds to the mean value of the squared redshift distribution $N^2(z)$. This is a natural choice because $\xi(r_p, \pi)$ depends on the distribution of galaxy pairs. The goodness of this choice has then been verified using the mock samples, where both the value of $\Omega_m(z)$ and $b_L(z)$ are known or can be directly recovered. In this way, the behaviour of $\beta(z) = \Omega_m^{0.55}(z)/b_L(z)$ can be directly compared to the estimated global value from the whole redshift range. This shows that our estimate of $\beta$ should coincide with $\beta(z = 0.77)$ within 3%, which is well below our statistical errors (see Supplementary Information).

This is the first measurement of $\beta$ at a redshift approaching unity based on a fully homogeneous galaxy redshift survey over a large volume and with accurate control over selection biases, finite sampling and cosmic variance errors. The detection and quantitative measurement of galaxy streaming motions at an epoch when the Universe was significantly younger is an important observational result in itself, testifying to the gradual growth of structure and corroborating the gravitational instability picture. To translate this measurement into an estimate of the growth rate $f = \beta b_L$, we need to know the effective linear bias factor characterizing the mean relative clustering of our galaxies with respect to the underlying mass. With sufficient statistics, $b_L$ can be determined directly from the survey data, by measuring the higher-order details of the clustering pattern[20], but this would require a survey several times larger than that used here.

We thus adopt a different approach that has already been successfully applied to the Deep part of the VVDS survey[23]. This requires including additional information provided by independent observations, such as the level of anisotropy in the Cosmic Microwave Background[24] or the mean number density of galaxy clusters[25,26]. Both of these measurements constrain the root-mean-square amplitude of mass density fluctuations on a given scale, conventionally measured in spheres of $8h^{-1}$ Mpc radius and indicated as $\sigma_8$. The Cosmic Microwave Background data from the Wilkinson Microwave Anisotropy Probe (WMAP) experiment[24] indicate that $\sigma_8^{mass}(z = 0) = 0.78 \pm 0.03$. This allows us to estimate $b_L = \sigma_8^{gal}(z = 0.77)/\sigma_8^{mass}(z = 0.77)$. $\sigma_8^{gal}(z = 0.77)$ is measured directly from the sample by counting the number of galaxies in randomly placed spheres; the corresponding mass value is instead obtained by scaling the WMAP value to $z = 0.77$ using linear theory in a self-consistent cosmology (which has a weak influence on the final result, for a flat geometry). In this way, we obtain $b_L = 1.3 \pm 0.1$, corresponding to a growth rate of $f(z = 0.77) = \beta b_L = 0.91 \pm 0.36$.

It is interesting to compare this measurement to available model predictions (Fig. 2). These include the standard flat ($\Omega_{m0} = 0.25$, $\Omega_{\Lambda 0} = 0.75$) cosmological-constant model, an open model with the same $\Omega_{m0}$ but no cosmological constant $\Lambda$, the DGP braneworld modification of gravitational theory[7] and two cases in which the dark matter component interacts with the dark energy field[5]. Clearly, error bars on this measurement alone are still too large to discriminate among these models. We also show in Fig. 2 the few existing measurements of $f$ at lower redshift. These include a value at $z \approx 0.15$ from the 2dFGRS[21] and another estimate at $z = 0.55$ that we have computed using a recent measurement of $\beta$ from a survey of luminous red galaxies[27]. This value can only be taken as indicative, as it was obtained by an analysis that tries to account for extra distortions due to the geometric Alcock–Paczynski effect[28] and imposes the additional constraint of $\Omega_m$ matching the evolution of clustering to $z \approx 0$ (see Fig. 2 caption and Supplementary Information).

With these caveats in mind, it is nevertheless encouraging to observe a coherent trend in the measurements. In particular, considering the standard general theory of relativity framework, even with the current error bars the evolution of the growth rate evidently disfavours a Universe with open geometry containing only matter (at the level of ~25% of the critical density as measured by several independent probes[10,21,25,26]). This is a relevant result, as it represents an indication, independent of the Cosmic Microwave Background[24], of the need of extra dark energy to bring the curvature close to zero. We note that a purely illustrative $\chi^2$ fit of the three data points to the functional form $f(z) = [\Omega_m(z)]^\gamma$ would indeed favour the flat cosmological constant model with growth index $\gamma \approx 0.55$–0.6, although with rather low confidence.

To discriminate among different dark energy or modified-gravity models at a finer level will require more precise estimates of $\beta$ and $b_L$ or a larger number of independent measurements with similar precision. Ongoing and planned redshift surveys are expected to fulfil this need in the near future, both in quantity and quality (see Supplementary Information). Overall, these results suggest that redshift-space distortions will become a primary method in the quest to identify the nature of cosmic acceleration.

**Supplementary Information** is linked to the online version of the paper at www.nature.com/nature.

**Acknowledgements** LG thanks M. Longair, C. Baugh, C. Porciani, P. Norberg, J. Peacock, A. Szalay and Y. Wang for discussions, S. White for suggestions and encouragement and L. Amendola, C. Di Porto and E. Linder for providing model predictions in electronic form. G. Pratt, S. White and E. Linder are gratefully acknowledged for reading versions of the manuscript. LG acknowledges support and hospitality of MPE, MPA and ESO during the development of this work. This research has been developed within the framework of the VVDS consortium and has been partially supported by the CNRS-INSU and its Programme National de Cosmologie (France), and by PRIN-INAF 2005. The VLT-VIMOS observations were carried out on guaranteed time allocated by the European Southern Observatory (ESO) to the VIRMOS consortium, under a contractual agreement between the CNRS of France, heading a consortium of French and Italian institutes, and ESO, to design, manufacture and test the VIMOS instrument. This paper is dedicated to the memory of Peter Schuecker, beloved colleague and friend.

**Author Information** Reprints and permissions information is available at www.nature.com/reprints. Correspondence and requests for materials should be addressed to L.G. (luigi.guzzo@brera.inaf.it).

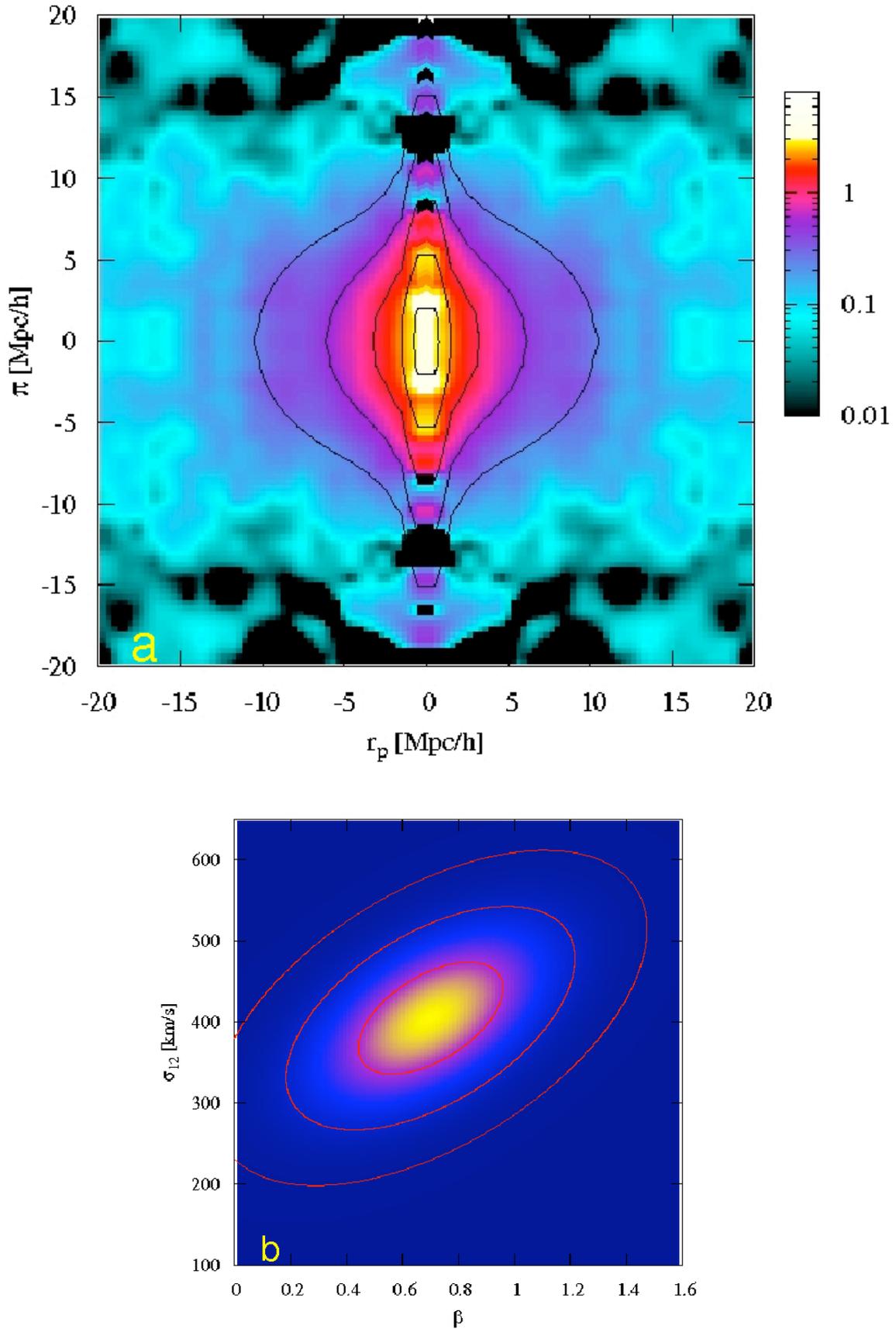

**Figure 1** **Estimate of the degree of distortion induced by coherent motions on the measured large-scale distribution of galaxies at high redshift.** For a given mean density of matter, this depends on the amount of dark energy and is quantified by the level

of anisotropy in the galaxy correlation function $\xi(r_p,\pi)$. **a**, Colour-coded representation of $\xi(r_p,\pi)$ as measured using ~6,000 galaxy redshifts with $0.6 < z < 1.2$ (effective redshift $\langle z \rangle = 0.77$) in the VVDS-Wide survey. The intensity describe the measured degree of correlation as a function of the transverse ($r_p$) and radial ($\pi$) separation of galaxy pairs. $\xi(r_p,\pi)$ has been computed in pixels of $1h^{-1}$ Mpc per side and smoothed with a gaussian kernel before plotting. The actual measurement is replicated over four quadrants to show the deviations from circular symmetry. Galaxy peculiar velocities combine with the cosmological expansion, producing the distorted pattern when the redshift is used as a distance measure. In the absence of peculiar motions, the contours would be perfect circles. The effect of galaxy infall caused by the growth of large-scale structure is evident in the flattening of the blue–green large-scale levels, while the small-scale elongation along $\pi$ (white–yellow–red contours) is the result of the high-velocity-dispersion pairs in group and clusters of galaxies ('fingers of God'). The superimposed solid contours correspond to the best-fitting distortion model with a compression parameter $\beta = 0.70$ and a pairwise dispersion $\sigma_{12} = 412$ km s$^{-1}$, obtained by maximizing the model likelihood given the data. **b**, Confidence levels for the compression parameter $\beta$ and the dispersion of relative velocities of galaxy pairs, $\sigma_{12}$; the contours correspond to the bivariate gaussian that best reproduces the distribution of 100 Monte Carlo measurements on fully realistic mock realizations of the data, constructed from numerical simulations (see Supplementary Information). The solid lines correspond to one-parameter confidence levels of 68%, 95% and 99%, such that marginalizing over $\sigma_{12}$, we obtain the root-mean-square uncertainty on $\beta$.

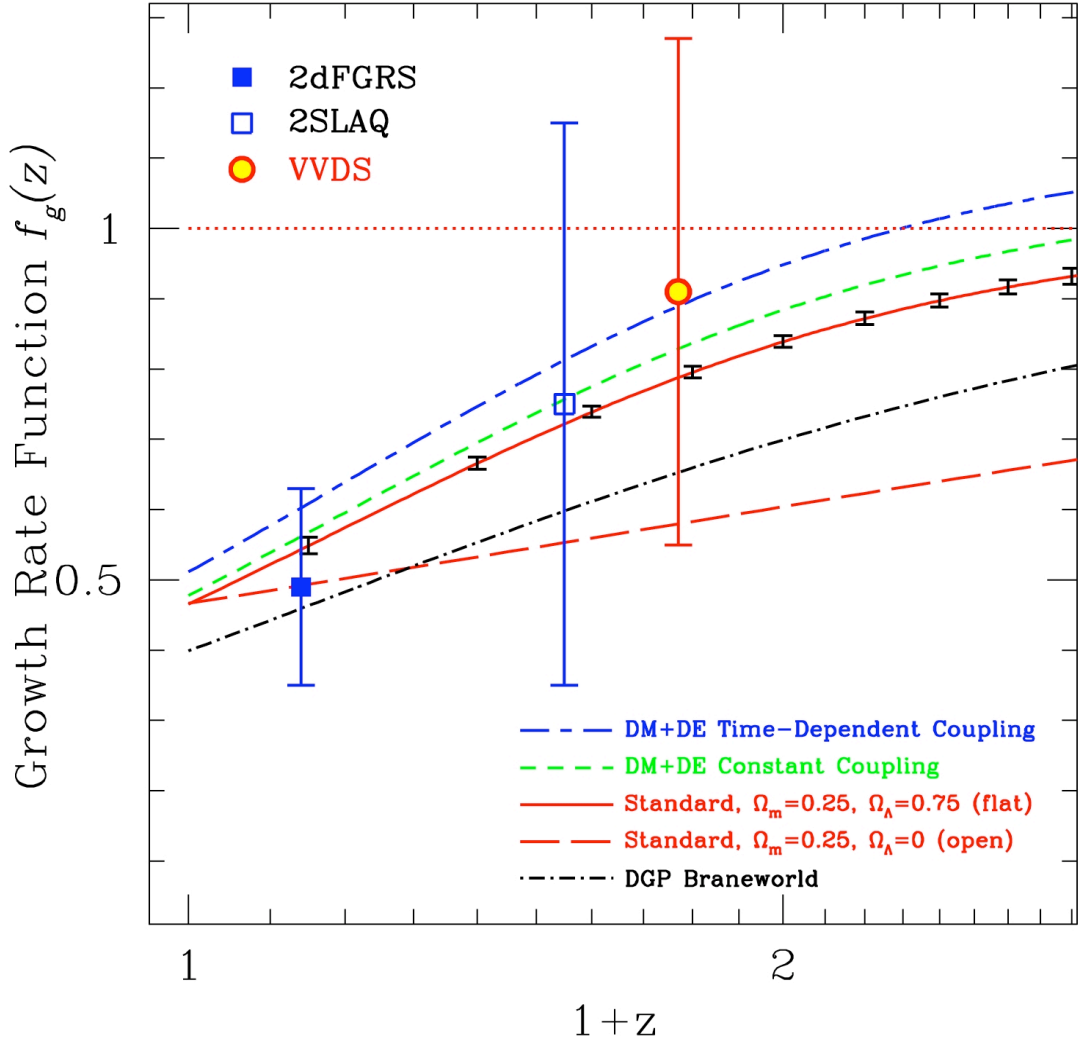

**Figure 2 Estimates of the growth rate of cosmic structure compared to predictions from various theoretical models**. Values of $f = \beta b_L$ are plotted as a function of the inverse of the cosmic expansion factor $1 + z = a(t)^{-1}$. Our new measurement at $z = 0.77$ from the VVDS-Wide survey (red circle) is shown together with that from the 2dFGRS, computed from the published[21] value of $\beta$; to do this, we adopted the bias value $b_L = 1.0 \pm 0.1$ estimated from higher-order clustering in the same survey[20]. We have also used very recent measurements from the 2dF-SDSS LRG and QSO (2SLAQ) survey of luminous red galaxies[27] (blue open square) to add one further point at $z = 0.55$. In this case, however, the values of $\beta$ and $b_L$ are not fully independent, because they have been obtained by imposing simultaneous consistency with the clustering measured at $z = 0$. In practice, this forces the resulting $f$ towards the flat $\Lambda$ model, that is, $\sim \Omega_m^{0.55}$. A more appropriate treatment would require an independent estimate of the bias for this sample[23]; this uncertainty is accounted for by the error bars, which in all cases correspond to 68% confidence intervals. The solid red line gives the growth rate for the standard cosmological-constant flat ($\Omega_{m0} = 0.25$, $\Omega_{\Lambda 0} = 0.75$) model, while the dashed red line is the corresponding open model with the same matter density but no cosmological constant; the blue and green dashed curves describe models in which dark energy is coupled to dark matter[5]; the black dot-dashed line is the DGP braneworld model, an extra-dimensional

modification of the gravitation theory[7]. For reference, the red horizontal dotted line $f(z) \equiv 1$ corresponds to the constant growth rate we expect in a critical-density Einstein–De Sitter model, in which the flat geometry is due to matter only ($\Omega_{m0} = 1$). Interestingly, despite the large error bars, the available measurements coherently indicate the need for a low $\Omega_m$, but at the same time disfavour an open model, thus requiring the presence of a cosmological constant or dark energy. We also provide an example of the accuracy achievable by future surveys in discriminating which kind of dark energy model is correct: the small black error bars on the standard $\Lambda$ curve (red) show forecasts for measurements in bins of size $\Delta z = 0.2$ from an all-sky survey of a half-billion infrared-selected ($H < 23$) galaxies, as recently proposed to the ESA Cosmic Vision programme by the SPACE

# Supplementary Information: A test of the nature of cosmic acceleration using galaxy redshift distortions

## THE STANDARD MODEL OF COSMOLOGY

An unprecedented convergence of observational results over the last few years indicates that we live in a low-density, expanding Universe with spatially flat geometry, that, quite surprisingly, appears to have recently entered a phase of accelerated expansion. This latter evidence emerged consistently from independent observations of the light dimming in distant supernovae, used as "standard candles" to probe the expansion history of the Universe[1,2]. At the same time, the spectrum of anisotropies in the Cosmic Microwave Background (CMB)[3] implies to high accuracy that the metric of the Universe as a whole is Euclidean, i.e. corresponding to an effective cosmic density parameter $\Omega_{TOT}=1$ [defined, at any epoch, as the ratio of the total average density $\rho_{TOT}$ to the "critical" density $\rho_c = 3H(t)^2/8\pi G$, where $H(t)$ is the value of the Hubble parameter $H(t) = d\ln a(t)/dt$, $G$ is Newton's gravitational constant and $a(t)$ is the cosmic expansion factor]. Several direct determinations of the contribution of matter, $\Omega_m$, to the global mass-energy budget indicate, on the other hand, that $\Omega_m \approx 0.25$, with only ~4% of the cosmic density provided by conventional "baryonic" matter, and ~21% in the form of a "dark matter" of unknown nature[4,5]. The flat geometry indicated by the CMB anisotropy spectrum therefore requires that "something else", neither conventional nor dark matter, must be providing the missing mass-energy needed to have $\Omega_{TOT}=1$. Remarkably, although its nature is completely mysterious, the physical entity producing the acceleration deduced by the observations of distant supernovae is capable of filling this gap, providing a "dark energy" density parameter $\Omega_\Lambda \approx 0.75$. Current observations are compatible with this extra contribution corresponding simply to a non-zero *cosmological constant* in the equations of General Relativity (GR). This term was originally introduced by A. Einstein to obtain a static solution when applying his theory to the Universe as a whole, but soon discarded after E. Hubble's discovery of universal expansion. The current evidence for cosmic acceleration emerges from measurements of the Hubble expansion parameter $H(z)$ [where the redshift $z(t) = a^{-1}(t) - 1$], that for a flat geometry can be written as

$$H^2(z) = H_0^2\left[\Omega_{m_0}\left(1+z\right)^3 + \Omega_{\Lambda_0} X(z)\right] , \qquad (S1)$$

with

$$X(z) = \exp\left\{3\int_0^z (1+z)^{-1}[1+w_X(z)]dz\right\} \quad . \tag{S2}$$

Here $X(z)$ corresponds to an extra component of "dark energy" with generic equation of state $w_x(z) = p_x(z)/c^2\rho(z)$; the cosmological constant case corresponds to $w_X(z) \equiv -1$, but any fluid with $w_X(z) < -1/3$ would also lead to an accelerated expansion. High-redshift Type Ia supernovae used as standard candles probe $H(z)$ via their cosmological luminosity distance[1,2]. Similarly, proposed new experiments based on detecting the imprint of *Baryonic Acoustic Oscillations (BAO)* in the clustering of galaxies [6,7], use the typical scale of BAO as a "standard rod", probing $H(z)$ through the angular-diameter distance[8].

## GROWTH OF DENSITY PERTURBATIONS

In the limit of small perturbations, the equations describing the evolution of density fluctuations in an expanding Universe can be linearized, obtaining the well-known growth equation [9,10]

$$\ddot{\delta} + 2H(t)\dot{\delta} = 4\pi G\langle\rho\rangle\delta \quad , \tag{S3}$$

where dots indicate time derivatives, $H(t)$ is the cosmic expansion parameter and $\delta(\vec{x},t) = (\rho(\vec{x},t) - \langle\rho(t)\rangle)/\langle\rho(t)\rangle$. This differential equation has a growing solution $\delta^+(\vec{x},t) = \hat{\delta}(\vec{x})D(t)$ (ref. 11). The equation of mass conservation, if the mass distribution is modelled as a continuous pressureless fluid, is

$$\frac{\partial\delta}{\partial t} + \frac{1}{a}\nabla\bullet(1+\delta)\vec{v} = 0 \quad , \tag{S4}$$

which, inserting the growing solution $\hat{\delta}(\vec{x})D(t)$ yields a peculiar velocity field

$$\vec{v}(\vec{x}) = \frac{fH_0 a}{4\pi}\int_V \frac{\vec{y}-\vec{x}}{|\vec{y}-\vec{x}|^3}\delta(\vec{y})d^3y \quad , \tag{S5}$$

where the factor

$$f = \frac{d\ln D}{d\ln a} \tag{S6}$$

is the *linear growth rate* of fluctuations. The growth rate essentially depends on the value of the cosmic matter density at the given epoch $\Omega_m(z)$ (refs. 11, 15). In fact, it has been shown that for a wide range of dark energy and modified gravity models, one can write to very good accuracy that $f \approx \Omega_m^\gamma$, where $\gamma = 0.55 + 0.05[1 + w(z=1)]$ and $w$ is the effective equation of state. This leads (e.g. Figure 2, main paper), to $\gamma = 0.55$ for the cosmological constant model in a GR background and $\gamma = 0.68$ for the DGP "braneworld" model (see refs. 15 and 16 and further in main paper). The growth rate is directly related to the linear redshift distortion parameter we measure in this work as $\beta = f / b_L$, where $b_L$ is the *linear bias value*, that can be defined as the ratio between the root-mean-squared density contrasts in the galaxy and mass distributions on scales R where linear theory applies: $b_L = \sigma_R^{gal} / \sigma_R^{mass}$ .

### THE VLT-VIMOS DEEP SURVEY

The VIMOS-VLT Deep Survey (VVDS) was designed to probe the combined evolution of galaxies and large scale structure to z~2, (reaching up to z ~ 4.5 for the most extreme objects[17]), measuring of the order of 100,000 faint galaxy redshifts. The VVDS is built around the VIMOS multi-object spectrograph at the ESO VLT, capable of collecting simultaneously up to ~600 spectra[18]. The survey is composed of two distinct parts with complementary science goals: VVDS-Deep, currently covering 0.5 deg$^2$ to an apparent red magnitude $I_{AB}$ = 24 (5-hour exposures), focused on studying galaxy evolution and clustering on relatively small scale, and VVDS-Wide, covering ~7 deg$^2$ to $I_{AB}$ = 22.5 (1-hour exposures), focused on measuring galaxy clustering at z~1 on scales approaching ~100 h$^{-1}$ Mpc [19]. Virtually all results published so far are based on the "First Epoch" set of 6530 reliable (> 80% confidence) redshifts from the "F02" field of VVDS-*Deep*. The shallower *Wide* survey is ongoing and has so far collected ~30,000 spectra. The results presented in this paper are based on the first complete subset of VVDS-Wide redshifts, centred on the F22 Wide field which covers 4 deg$^2$ and includes 11,400 galaxy redshifts between z=0 and z~1.3, spanning a total volume of 7.9 x 10$^6$ h$^{-3}$ Mpc$^3$. Both the VVDS-Deep and VVDS-Wide spectra were collected during the guaranteed-time observations awarded to the VVDS Consortium for the construction of VIMOS.

## REDSHIFT- AND REAL-SPACE CORRELATION FUNCTIONS

The simplest statistic for studying the inhomogeneity of the galaxy distribution is the two-point correlation function $\xi(s)$. This measures the excess probability above random of finding a pair of galaxies with separation $|\vec{s}|$. Galaxy distances are proportional to the observed *red-shift* of emission/absorption lines in their spectra with respect to their laboratory value. This provides us with a mean to re-construct the 3D distribution of galaxies, and thus measure their relative separations. However, galaxy peculiar velocities add a Doppler component to the cosmological redshift, thus shifting the apparent galaxy position with respect to its correct, real-space position. This distorts galaxy redshift maps and modifies the measured two-point correlations. To measure this peculiar velocity contribution, it is convenient to separate $\xi$ into a function of two variables, de-composing $\vec{s}$ as $\vec{s} = \vec{r}_p + \vec{\pi}$, where $r_p$ and $\pi$ are respectively perpendicular and parallel to the line-of-sight [20]. Peculiar motions will thus affect the variable $\pi$ only, and the resulting correlation map $\xi(r_p, \pi)$ will show distortions along the ordinate axis only[21]. In practice, $\xi(r_p, \pi)$ is estimated from the data by counting the number of galaxy pairs in bins of size $(\Delta r_p, \Delta \pi)$ over a grid of separations $(r_p, \pi)$, and comparing them to those from a random sample with the same geometry and selection function; the counts of galaxy and random pairs are combined through appropriate estimators that allow the inclusion of normalized weights to account for observational effects (see next section)[22]. Here we use the widely applied minimum-variance estimator of Landy and Szalay[23]. Once $\xi(r_p, \pi)$ has been estimated, one can then recover the undistorted real-space correlation function $\xi(r)$ by projecting $\xi(r_p, \pi)$ along the line of sight direction, constructing the *projected function*

$$w_p(r_p) \equiv 2 \int_0^\infty \xi(r_p, \pi) d\pi = 2 \int_{r_p}^\infty \frac{y \xi_r(y) dy}{\left(y^2 - r_p^2\right)} \quad , \tag{S7}$$

where in the last term $\xi_r(y)$ is the real-space 1-dimensional correlation function we are looking for, given the independence of the variable $r_p$ on the redshift distortions. This integral can be inverted numerically to recover $\xi_r$ (refs. 24, 25). The accuracy of this operation is crucial for the estimate of $\beta$, as $\xi_r$ describes the reference clustering value in the linear distortion model. We have tested directly on our mock surveys that the inversion of $w_p(r_p)$ provides the best estimate of the spatial correlation function, with respect to other methods [as assuming a power-law form or other approximations to $\xi_r(r)$], that tend to bias the final value of $\beta$ [25,26].

### *ESTIMATING DYNAMICAL QUANTITIES FROM REDSHIFT-SPACE DISTORTIONS*

The best way to extract quantitative dynamical information contained in the redshift-space correlation function $\xi(r_p,\pi)$, is to expand it using a base of spherical harmonics, a procedure commonly used in physics when dealing with problems that present a spherical or circular symmetry [27,28]

$$\xi_L(r_p,\pi) = \xi_0(s)P_0(\mu) + \xi_2(s)P_2(\mu) + \xi_4(s)P_4(\mu) \quad . \tag{S8}$$

Here $\mu = \hat{r} \bullet \hat{\pi}$ is the cosine of the angle between the separation vector and the line of sight, $P_n(\mu)$ is the Legendre Polynomial of order n and $\xi_n(s)$ the corresponding moment of $\xi(s)$. $\beta$ can be extracted via appropriate combinations of these moments, or through direct modelling of $\xi(r_p,\pi)$. The suffix $L$ of $\xi_L(r_p,\pi)$ indicates that this model includes only the linear distortions, i.e. those related to large-scale coherent motions. To be complete, the model for $\xi(r_p,\pi)$ needs to include the non-linear distortion due to high-velocity pairs in groups and clusters of galaxies, that on galaxy maps transforms these spherical agglomerates into cigar-like structures known as "Fingers of God". This effect is responsible for the elongation of the contours at small $r_p$ in Fig.1 of the main paper. This is modelled by convolving the linear $\xi_L(s)$ with the distribution function of 1-dimensional relative velocities of galaxy pairs along the line of sight $\varphi(v)$. Observations and numerical simulations indicate that this function is well described by a normalized exponential $\varphi(v) = \left(\sigma_{12}\sqrt{2}\right)^{-1}\exp\left\{-\sqrt{2}|v|\sigma_{12}^{-1}\right\}$. In this expression, $\sigma_{12}$ is the 1-dimensional pairwise velocity dispersion describing the strength of small-scale "thermal" random galaxy motions[29]. Including this contribution, the model for $\xi(r_p,\pi)$ is complete and becomes

$$\xi(r_p,\pi) = \int_{-\infty}^{\infty} \xi_L\left[r_p, \pi - \frac{v(1+z)}{H(z)}\right]\varphi(v)dv \quad , \tag{S9}$$

where $H(z)$ is the Hubble parameter (eq. 1) and the $(1+z)/H(z)$ factor properly accounts for the conversion of velocity shifts into comoving coordinate shifts at $z >> 0$. The model therefore depends on two free parameters, the linear compression $\beta$ and the velocity dispersion $\sigma_{12}$. We therefore define a likelihood function $\mathcal{L}$ of the model and the observed $\xi(r_p,\pi)$, as

$$-2\ln\mathcal{L} = \chi^2 = \sum_i \sum_j \frac{\left(y_{ij}^{(mod)} - y_{ij}^{(obs)}\right)^2}{\sigma_{ij}^2} \quad , \tag{S10}$$

where

$$y_{ij}^{(x)} = \log\left\{1 + \xi^{(x)}(r_{p_i}, \pi_j)\right\} \tag{S11}$$

and the errors $\sigma^2_{ij}$ are computed from the scatter of $y_{ij}$ among 100 mock samples (see below). The best-fit values for $\beta$ and $\sigma_{12}$ are then found by maximizing the likelihood of the model, given our measurements. This expression of the likelihood, involving the logarithm of the excess number of pairs in each bin $(1+\xi)$, has been shown to perform better than a direct likelihood fit on the values of $\xi$, as it reduces the weight of the small, non-linear scales where $\xi(r_p,\pi)$ has the largest values [30]. This is confirmed by our Monte Carlo experiments (see next section), that show that this estimator is unbiased at the level of accuracy reachable in this work. This is particularly important, as in applying eq. S10 we are implicitly assuming that the covariance matrix of the data is diagonal, i.e. that the values of $\xi(r_p,\pi)$ in different bins are independent, which is known not to be the case. Estimating properly the off-diagonal elements of the covariance matrix using mock samples or other techniques is neither straightforward nor computationally easy given the size the matrix has for a bi-dimensional quantity like $\xi(r_p,\pi)$. Our accurate mock samples allow us to overcome this difficulty in a pragmatic way, i.e. by testing directly whether our estimator is biased by this assumption. In fact, the mean value from the 100 mock samples – analysed in exactly the same way – turns out to be unbiased to better than 3%, which allows us to conclude that the effect of non-diagonal terms on the estimate of $\beta$ is negligible, at least at this level.

The same tests also indicate that only bins with $r_p < 20$ $h^{-1}$ Mpc and $\pi < 20$ $h^{-1}$ Mpc can be conveniently used in the fit, to recover $\beta$ with minimum bias. Above these scales $\xi(r_p,\pi)$ is dominated by the noise. Following previous analyses [30], we have also checked the effect of excluding from the fit the strongly non-linear scales at small separations. Our estimate turns out to be rather insensitive to inclusion or exclusion of the bins with $r_p < 3$ $h^{-1}$ Mpc, with a tendency to bias $\beta$ slightly high when these are excluded. This limited sensitivity to the non-linear contribution is related to our observing strategy, which significantly under-samples the small scale high-velocity pairs. Typical of large redshift surveys performed with multi-slit or fiber-optic spectrographs, VIMOS observations tend to be biased against very close pairs on the sky: galaxies closer than ~10 arcsec cannot both be measured within the same observation, due to the physical size of the slits on the VIMOS multi-object mask. The effect on the measured clustering can be corrected statistically in an effective way through the inclusion of appropriate weights in the correlation function estimator[22,26] . After experimenting directly with the mock samples – which precisely include the same under-counting of small pairs, having been "observed" through the same SSPOC software used to prepare the observations – we decided not to apply here any of our usual corrections, apart from the obvious ones accounting for the global selection function. Our thoroughly tested VVDS weighting scheme [22] was designed to recover missing clustering on scales <2 $h^{-1}$ Mpc and is relevant if one is interested in measuring the shape of $\xi(r)$ at these separations. However, this also has the consequence of *enhancing* the non-linear contribution to $\xi(r_p,\pi)$ (i.e. that due to very close pairs in galaxy clusters), which for this analysis is only a nuisance. In fact, our tests show that under-sampling of close pairs (particularly effective in the F22-Wide data used here, for which VIMOS visited only once every point of the field), has in fact a beneficial effect on the stability of our measurement of $\beta$ by artificially reducing the statistical weight of dense, virialized regions. This implies that the measured value of the pairwise dispersion $\sigma_{12}$ is certainly underestimated, but this is not a problem for this analysis. Rather, this "forced sparse-sampled" strategy, which under-weights the strongly non-linear scales, appears to be particularly appropriate for measuring $\beta$ using the convolution model (eq. S8), in which the treatment of non-linear distortions is inevitably an approximation [29]. It should also be noted, however, that the value of the pairwise dispersion $\sigma_{12}$ we obtain by applying the same procedure to the Millennium mock catalogues is very close to that we obtain from the data. This implies that, at least after filtering on ~2 $h^{-1}$ Mpc scales, the simulations provide a very good description of the observed dynamics of galaxies at z~1, both in their linear and non-linear components.

## MONTE CARLO ESTIMATE OF ERRORS AND STATISTICAL SIGNIFICANCE

The error quoted on our measurement of $\beta$ has been estimated from the scatter within a set of 100 fully realistic mock VVDS surveys. These were constructed by "observing", under exactly the same conditions as the real data, a set of simulated galaxy catalogues obtained by applying a semi-analytic model of galaxy formation [31] to the dark-matter halos identified in "light-cones" from the Millennium Simulation [32]. This combination of fairly large simulation size (each time step is a box of 500 h$^{-1}$ Mpc side), mass resolution and accurate semi-analytic modelling represents the current state-of-the-art for this type of simulated surveys; these specific models have been shown to correctly reproduce, at least to z~1.2, several key observational properties of galaxies, such as their number counts, luminosity function and redshift distribution [33]. It is reasonable to think, therefore, that our error estimates include both statistical errors and *cosmic variance* (i.e. the scatter in the measurements expected if N other samples of the same size were observed and analyzed). Rigorously speaking, given the total volume of the Millennium simulation and that of our survey, the number of independent mock samples that can be accommodated within the volume of the computational box is 21. However, the probabilistic assignment of semi-analytic galaxies to dark-matter halos identified in the simulation and the simulated flux-limited "observation" add further variance to each mock sample. Different factors contribute to this. For example, even if the same portion of simulation volume is included in different light-cones, the corresponding dark-matter halos might or might not correspond to galaxies in the final mock surveys. This will in fact depend on the relative distance to the observer and thus on the apparent magnitude of the object; if this is fainter than the survey flux limit, it will be excluded from the sample. Another even more important source of variance among the mock catalogues for this specific analysis is due to the very nature of our measurement: the same volume of the simulation will give completely independent linear distortions (and thus a different estimate of $\beta$), depending on the direction under which it is observed. This increases *de facto* by a factor of 3 the degrees of freedom on $\beta$ for mocks extracted from a fixed volume, given that the light cones are constructed by stacking randomly oriented simulation boxes[37]. Taken together, these factors more than justify our claim that the 100 simulated catalogues used here can be considered as substantially independent realizations of our survey.

These Monte Carlo experiments allow us also to assess the contribution of observational systematic effects and quantify any bias in our modelling of redshift distortions [25,26]. The main results from this exercise are shown in Supplementary Figure 1. We find that our estimator of $\beta$ is substantially unbiased within the statistical error, with a mean expectation value from 100 mocks of $\beta$=0.62±0.03. This is well within one standard deviation of the "true" value of the parent Millennium simulation, $\beta$=0.64±0.03 (where the latter error comes entirely from the estimate of galaxy bias in the simulation). This reference value is obtained directly from the relation

$$\beta(z) = \Omega_m^{0.55}(z)/b_L(z), \quad \text{with} \quad \Omega_m(z) = \left\{1 + (1 - \Omega_{m,0})/\left[\Omega_{m,0}(1+z)^3\right]\right\}^{-1}, \quad \text{knowing that}$$

$\Omega_{m,0}$=0.25, and estimating the bias directly as $b_L = \sigma_8^{gal}(z=0.77)/\sigma_8^{mass}(z=0.77)$, where both *rms* fluctuations in galaxies and mass are clearly known *a priori* in the case of the simulation. Interestingly, we also find that a significant reduction of the *rms* error (about a factor of 3), is obtained when in our model (eq. S8) we use the "true" spatial correlation function $\xi(r)$ (directly estimated in the case of the simulation), instead of the de-projected one from eq. S7. This information is obviously not available for the real data, but this exercise indicates that there is room for a significant gain in the accuracy of the estimate of $\beta$, if our knowledge of the underlying correlation function can be improved, e.g. via independent data or modelling.

## SELF-CONSISTENCY OF DISTANCE-REDSHIFT CONVERSIONS: THE ALCOCK-PACZYNSKI EFFECT

To perform the measurements presented here, the measured redshifts from our galaxy catalogue have been converted into co-moving coordinates assuming a "concordance" model with matter density $\Omega_m$=0.3 and time-independent ($W = -1$) dark energy density $\Omega_\Lambda$=0.7. This means assuming *a priori* a cosmological model, which on the other hand is what in the end one is trying to constrain through $\beta$. Adopting the wrong cosmology would induce a distortion on the resulting correlation maps that adds to the effect of peculiar velocities we aim to measure. This was first noted by Alcock & Paczynski [35] and proposed as a method to estimate the cosmological parameters. The problem is that this effect and the dynamical distortion produced by peculiar velocities are in principle super-imposed in the observed correlation function. A method to model both effects simultaneously has been proposed [36], but practical applications show a strong degeneracy between the parameters involved [37]. We adopted a more practical approach and tested our ability to separate out the Alcock & Paczynski effect using our extended set of simulations (which were built for a Universe with $\Omega_m=0.25$, $\Omega_\Lambda=0.75$). We re-computed co-moving distances assuming incorrect cosmologies corresponding to values of $\Omega_m(z=0)$ between 0.1 and 1 in steps of 0.1 (in a flat cosmology, $\Omega_\Lambda=1-\Omega_m$). Then, for this set of 10 samples with slightly different comoving distances, we re-computed $\xi(r_p,\pi)$ and estimated the corresponding $\beta$. In GR cosmology (as used to run the simulation) and knowing the bias, each value of $\beta$ yielded, in turn, a new value of $\Omega_m(z=0)$, which we used to re-compute distances, and so on. The encouraging (and somewhat surprising) result of this exercise is that $\Omega_m(z=0)$ rapidly converges to the true value of the simulation, whatever the starting value was. This is true independently of whether the initial value is larger or smaller than the true one. Thus, we conclude that our measurement of $\beta$ ($\Omega_m$) at z=0.77 is robust against the Alcock-Paczynski distortion. At the same time, this seems to imply that it is difficult in practice to use the Alcock-Paczynski effect to extract cosmological parameters from $\xi(r_p,\pi)$, at least at these redshifts, the dynamical effect of peculiar velocities being dominant [37].

## PROSPECTS FOR ONGOING AND FUTURE REDSHIFT SURVEYS

The method proposed here to map the growth rate as a function of redshift has a few specific advantages and will be fully exploited by the next generation of deep/wide redshift surveys. One advantage is that galaxies are used simply as test particles to probe a velocity field that depends on the mass distributed well outside the survey volume; for a given survey size, this makes the technique less sensitive to incompleteness or non-trivial selection effects. These have, for example, to be kept strictly under control when trying to measure the shape of the power spectrum of galaxy fluctuations on scales where the signal is just a few percent of the mean density (e.g. for measuring BAO). Secondly, it can be applied to relatively low-resolution spectroscopic surveys (requiring smaller amounts of telescope time), since it depends on a bulk effect and does not require a high precision in the measurement of each galaxy redshift. There are several ongoing redshift surveys that can be expected to be able shortly to provide independent measurements of $f(z)$ at different redshifts, with accuracy similar to the VVDS-F22 estimate of this paper.  These data sets include at high redshift the full VVDS-Wide survey, for which another ~4 deg² will soon become available, the similar Z-Cosmos survey [38] over 2 deg² and the deeper DEEP2 survey [39] over 3 deg².  Additionally, estimates of $\beta$ from the clustering of colour-selected QSO's have been recently obtained at z~1.4 using the same technique (and with the same drawbacks) of the red galaxies of Fig. 2 in the main paper [40], but need to be complemented with a more robust estimate of the bias to yield a reliable value of $f$.  Finally, at z~0 a second reliable and fully independent measurement should be within easy reach using the nearly one million redshifts of the Sloan Digital Sky Survey [41].

Looking more to the distant future, we have used our suite of mock surveys to produce some forecasts for the errors on β achievable by future redshift surveys with characteristics similar to VVDS Wide, but covering larger areas (Suppl. Fig. 2).  One finds that enlarging the survey volume (while keeping the galaxy density fixed), the *rms* error on $\beta$ scales as the square root of the volume.  The gain is slightly less effective if the density of sampling (i.e. the fraction of galaxies that are observed among all those brighter than the survey limit) is increased by the same factor. These dependencies can be combined into a useful empirical expression that we calibrated with the simulation results, as a function of the sampling rate $f_s$, the total mean density of objects expected <n> (in h³ Mpc⁻³) and its volume V (in h⁻³ Mpc³).  The relative error can be expressed as $\varepsilon_\beta \approx 50 / \left( \left( f_s \langle n \rangle \right)^{0.44} V^{0.5} \right)$. In the case of our z=[0.6-1.2] sample, the values $f_s$=0.2, <n>~5 x 10⁻³ h³ Mpc⁻³ and V=6.35 x 10⁶ h⁻³ Mpc³, correspond to the error estimated from the mocks.  Note that, interestingly, the total number of galaxies in the sample being $N = f_s \langle n \rangle Vol$, this formula indicates that the error scales almost as the square root of the number of galaxies.

These computations show that an extension of currently ongoing surveys will be able to push the statistical error on measurements of $\beta$ at z~1 below 10% (as e.g. in the case of the planned extension of VVDS-Wide to 16 deg²).  More ambitiously, a survey with the same depth (magnitude I<22.5) and sampling rate (~20%), but covering 100 times more area than F22 (400 deg²), would yield $\beta$ at z=0.8 to better than 4%.   Projects this size, however, will only be possible through new and dedicated instrumentation.  Several ideas are under discussion, including both ground-based and space-borne observatories.  Particularly attractive are the predictions on the accuracy achievable on $f(z)$ from an infrared-selected all-sky spectroscopic survey of ~10⁹ galaxies (the SPACE project [42]), reported for comparison as the very small error bars over the standard model in Fig. 2 of the main paper.   These errors assume precise knowledge of the galaxy bias factor, at the level expected from these larger surveys where it will be possible to apply higher-order statistics to measure it directly[43].   It is also very important to stress that measurements of the growth rate using redshift distortions will be possible from the same redshift surveys aiming at measuring the expansion history $H(z)$ using BAO, but with different systematics.   Additionally, it is expected that future large multi-band imaging surveys will provide estimates of the growth rate of mass fluctuations through gravitational lensing "tomography" [44].   Overall, therefore, an optimal strategy for a combined attack on the problem of explaining the cosmic acceleration over the next decade would seem to be that of having two complementary experiments, possibly both from space: an imaging survey capable of discovering large numbers of distant supernovae with accurate photometry, while making accurate weak lensing measurements, plus an infrared-selected all-sky spectroscopic survey to z~1.5-2, capable of measuring simultaneously to high precision both Baryonic Acoustic Oscillations (from the galaxy power spectrum) and the growth rate (using the technique proposed in this paper) within several redshift bins.

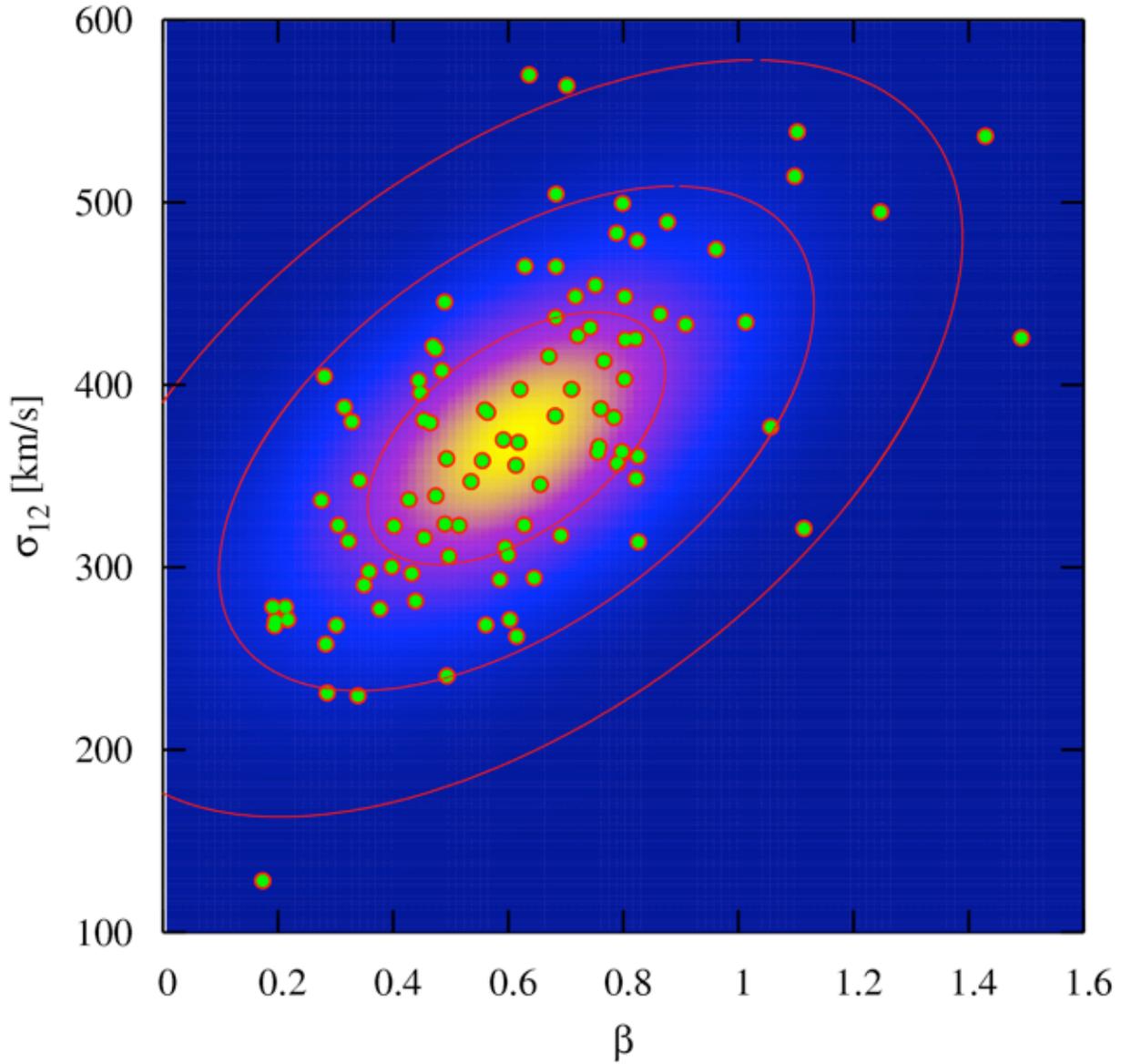

**Supplementary Figure 1. Estimates of the linear compression parameter $\beta$ and pairwise velocity dispersion $\sigma_{12}$ from 100 VVDS-like mock surveys.** Each measurement has been obtained by applying our model fitting procedure to mock samples that accurately reproduce the selection function and statistical properties of the actual F22 VVDS-Wide field. These have been constructed as "light-cones" [34] from the state-of-the-art Millennium Simulation [32], by applying a semi-analytic model of galaxy formation that correctly reproduces a large number of observed properties of galaxies and of their spatial distribution [31,33] . With this same set of mock surveys, we have also thoroughly explored the effectiveness of other methods of estimating $\beta$ from the moments of $\xi(r_p,\pi)$ (refs. 28, 30), concluding that for our data the direct fit using the de-projected spatial function gives the most stable and unbiased estimate. The colour scale gives the likelihood levels for the bi-variate Gaussian distribution that best describes the data. The red lines correspond to the standard loci for 1-parameter confidence levels of 68%, 95% and 99%, in the sense that their projection onto the two axes gives the corresponding intervals for the two parameters separately.

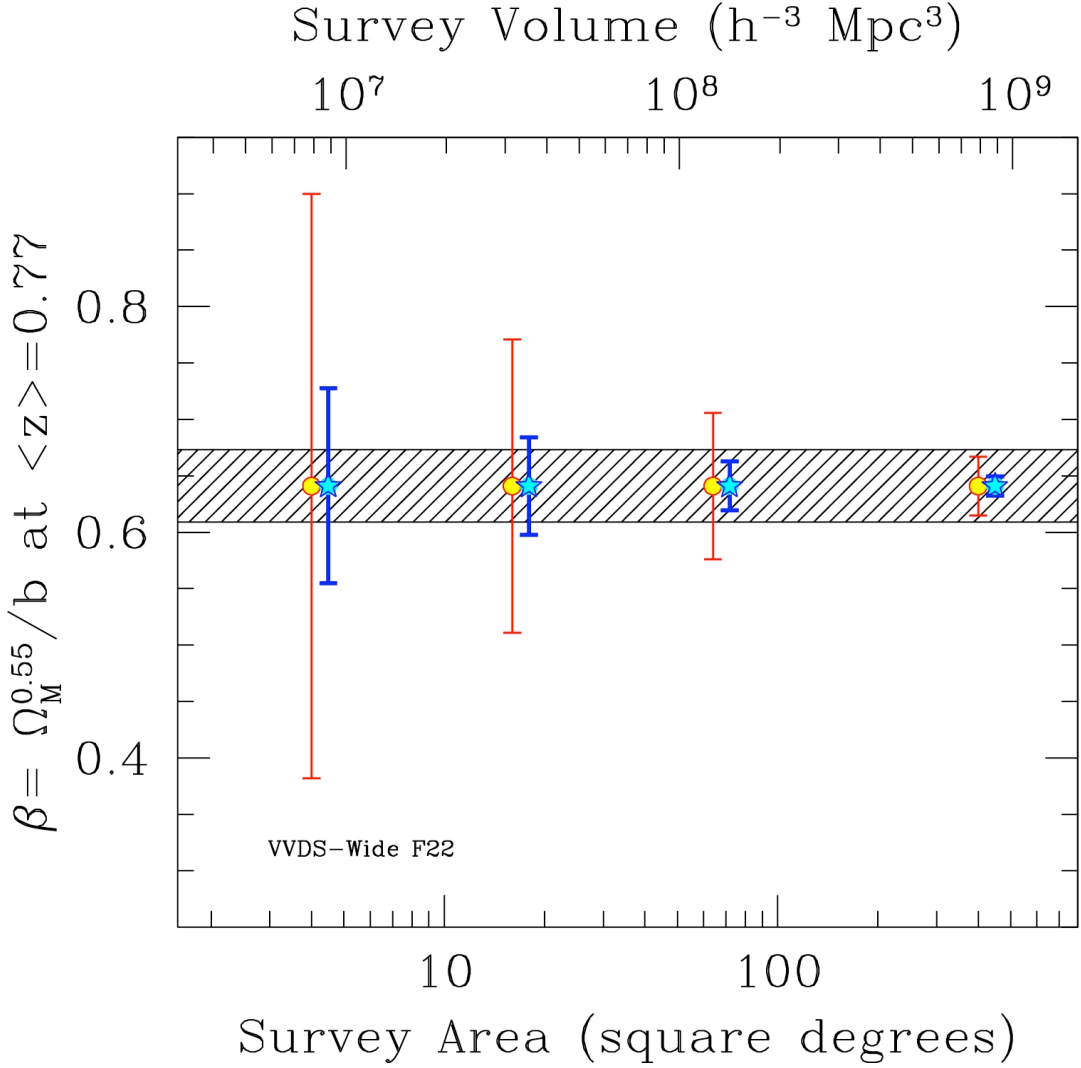

**Supplementary Figure 2. Dependence of the *rms* error of $\beta$ on the survey size.** The values have been obtained using mock surveys with a sampling of ~20% to a magnitude $I_{AB}=22.5$ and a redshift interval between z=0.6 and z=1.2, as for the VVDS survey data used in this paper. Points correspond to sky areas of 4, 16, 64 and 400 deg$^2$, respectively. The dashed band gives a ±5% error around the fiducial value of $\beta$ in the simulation. Red error bars correspond to using in the model for $\xi(r_p,\pi)$ (eq. S8) the real space function $\xi(r)$ obtained through de-projection of the observed $\xi(r_p,\pi)$ (eq. S7). Blue error bars show instead the ideal case in which the true real space function $\xi(r)$ (directly available in the case of the simulations) is used. This results in a factor of ~3 improvement, showing that a significant fraction of the error on $\beta$ is due to inaccuracies in our description of the intrinsic spatial clustering. Obviously, the true $\xi(r)$ is not known a priori for the data, but this result indicates that there is significant room for improving the measurement of $\beta$, also from current samples, provided we obtain a more precise estimate of $\xi(r)$. It should also be remarked that when statistical errors on $\beta$ approach the 3-5% level, as in the case of a 400 deg$^2$ survey, they become comparable to the estimated systematic errors. More detailed exploration of systematic effects and further optimization of the estimator would certainly be needed in this high-precision regime to understand how to reduce the errors even further.